\documentstyle[12pt]{article}

\input tcilatex

\begin{document}

\author{B.M. Levin\thanks{%
E-mail: {\it pl8m4004@peterlink.ru}}}
\title{Orthopositronium: `Annihilation of Positron in Gaseous Neon'}
\date{}
\maketitle

\vspace*{-15pt}
\begin{abstract}
On the basis of phenomenological model of the orthopositronium annihilation
`isotope anomaly' in gaseous neon (lifetime spectra, positrons source $^{22}$%
Na) the realistic estimation of an additinal mode ($\sim $0.2$\%$) of the
orthopositronium annihilation is received.

\end{abstract}


\thispagestyle{empty}
\vspace*{8pt}
\noindent In huge volume of the data on lifetime spectra of annihilation
in substance of positrons from $\beta ^{+}$-decay $^{\text{22}}$Na

\vspace*{-13pt}
$$
^{22}\text{Na}\,(3^{+})\to \,^{22*}\text{Ne}\,(2^{+})+e_\beta ^{+}+\nu , 
$$

\vspace*{1pt}
\noindent received for half a century of making of these measurements, the
experimental situation for gaseous neon proved to be {\it unique } \cite{bib1,bib2}.

In a method of delayed $\gamma _{\text{n}}$-$\gamma _{\text{a}}$%
-coincidences (`lifetime method') the appearance of positron in substance is
registered by nuclear $\gamma _{\text{n}}$-quantum (`start')%
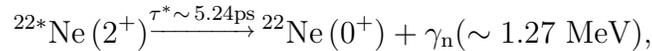
$$
^{22*}\text{Ne\,}(2^{+})^{\underrightarrow{\tau ^{*}\sim \,5.24\text{ps}}%
}\,^{22}\text{Ne\,}(0^{+})+\gamma _{\text{n}}(\sim 1.27\text{ MeV}), 
$$

\noindent and its annihilation --- on one of $\gamma _{\text{a}}$-quanta
(`stop') as a result of annihilation%
$$
e_\beta ^{+}+e^{-}\text{(in substance)}\rightarrow \text{most probably
annihilation on 2 or 3 }\gamma _{\text{a}}\text{-quanta.} 
$$
\noindent The counter, developing signal `stop', registers not only $\gamma
_{\text{a}}$-quantum (with energy $\leq $0.5 MeV) but also $\gamma _{\text{n}%
}$-quantum ($\sim $1.27 MeV) with recording efficiency 
$$
\varepsilon _{\text{0.5}}\cdot \stackrel{\_\_}{\Omega }_{\text{2}}\text{(}%
\leq \text{0.5)} 
$$

\vspace*{-9pt}
\noindent and

\vspace*{-17pt}
$$
\varepsilon _{\text{1.27}}\cdot \Omega _{\text{2}}\text{(1.27)} 
$$

\vspace*{3pt}
\noindent respectively, and the relation of random ($R$) and true ($C$)
coincidences looks like \cite{bib2}

$$
\frac RC=Q\Delta \tau \left[ 2+\frac{\varepsilon _{1.27}\cdot \Omega _2(1.27)%
}{\varepsilon _{0.5}\cdot \overline{\Omega }_2(\leq 0.5)}\right] , 
$$

\noindent where $Q$ (s$^{-1}$) is the power of the positron source, $\Delta
\tau $ is the resolving time of the coincidence circuit, $\varepsilon _{%
\text{i}}$ the recording efficiency for quanta entering the counter, and $%
\overline{\Omega }_2$ is the mean solid viewing angle of counter in
conditions of `bad' geometry.

The annihilation $\gamma _{\text{a}}$-quanta are registered in conditions of
`bad' geometry, as the annihilation acts occurs in all volume of gas, unlike
nuclear $\gamma _{\text{n}}$-quanta outgoing from the centre of volume,
where the source of positrons is lokated.

As the rate of occurrence of the marking `start' $\gamma _{\text{n}}$%
-quantum (1/$\tau ^{*}$) is more than the order higher than the rate of the
fastes processes of positrons annihilation in substance, the correctness and
accuracy of the lifetime method are not to be doubted.

Nevertheless once this common rule was broken in our article \cite{bib2} (its title
is repeated in the title of this paper). In this article the results of a
number of measurements executed in the USA, England, Canada and in Russia
(Institute of Chemical Physics, Russian Academy of Sciences, Moscow) are
generalized, and for the first time the attention is paid to specific
features of the positron annihilation lifetime spectra in the area of the so
called `shoulder' in gaseous neon as compared with other inert gases.
Collectivization of nuclear excitation $^{22*}$Ne$\,(2^{+})$ was assumed by us
in conditions of experiment in the macroscopic group of identical nuclei ($%
^{22}$Ne) --- like the nuclear gamma-resonance ({\it NGR}) --- as in the
natural mixture of the neon isotopes ($^{20}$Ne$,^{21}$Ne$,^{22}$Ne)
contains about 9\% $^{22}$Ne and then in such collective of nuclei it is
impossible to specify a location of a nucleus, which emits $\gamma _{\text{n}%
}$-quantum registered by the counter (`start'). On this basis the
explanation of the phenomenon of smoothing the shoulder in neon was offered,
which according to this version is caused by uncontrollable grows of a
background of random coincidance because of realization in neon of a
condition of `bad' geometry also for nuclear $\gamma _{\text{n}}$-quantum.
The paradox of this assumption, so long as the manifestation of {\it NGR} is
supposed to occur in {\it gas} (similar scattering of $\gamma $-quanta,
without recoil of the radiating and absorbing nuclei, is possible only in 
{\it a solid state} --- the M\"ossbauer effect), was attributed to
peculiarity of a final state of $\beta ^{+}$-decay of nucleus $^{22}$Na as
far back as on the initial stage of a study of the phenomenon of smoothing
of a shoulder in neon \cite{bib3}. We would like to notice that a year ago, on the
occasion of the fiftieth anniversary of the hypothesis of W.Pauli about
neutrino, B. Pontecorvo has marked: `\dots huge grows of neutrino physics,
which became a quantitative science, healthy and powerful and nevertheless
permit a quantitative unexpectedness' \cite{bib4}. Development of this thesis in
works \cite{bib5,bib6}, in view of the newest experimental data and the other technique
of `start' registration, based on the initial site of track of positron from 
$\beta ^{+}$-decay $^{22}$Na or $^{68}$Ga, allowed us to attribute formation
of {\it macroscopic collective /resonance/ nuclear state} ({\it MCN/R/S}) to
the specific quality of $\beta ^{+}$-decay of nuclei of this type $\Delta $J$%
^\pi =1^{+}$ (with a change of a spin on $\pm 1$ and conversation of
parity), which should be considered as {\it topological quantum transition}
\cite{bib5,bib7} in limited `volume' of space-time.

Earlier, as a result of taking into account all manifestations of positron
annihilation specific features in gaseous neon, it became clear, that the
shoulder in neon can be influenced also by annihilation of orthopositronium
(spin {\it S}$=1$, triplet state $^3(e_{\beta \,}^{+}e^{-})_1$, symbol $^{%
\text{T}}$Ps, $\tau _{\text{T}}\simeq 140$ ns) \cite{bib8}, while short living
parapositronium (spin {\it S}$=0$, singlet state $^1(e_{\beta
\,}^{+}e^{-})_0 $, symbol $^{\text{S}}$Ps, $\tau _{\text{S}}\simeq 125$ ps)
can not exert such influence, as it is registered by the device with
resolving time $\Delta \tau \sim 0.1\div 1$ns near interval of time $\Delta
t\sim 0$ (`peak of instant coincidence'). On this basis was constructed
phenomenological model of this phenomenon and parametrization of lifetime
spectra of positron (orthopositronium) annihilation in gaseous neon was
received \cite{bib3}.

This point of view received additional stimuli in the results of precise
measurements of the orthopositronium annihilation rate ($\lambda _{\text{T}%
}=1/\tau _T$), executed by a group of Michigan University (Ann Arbor, USA ---
A. Rich, D.W. Gidley, and co-workers) \cite{bib9}, which were confirmed and specified
in the second half of 1980s with use of two different techniques of
`start' registration \cite{bib10,bib11,bib12}. In these works the excess of experimental
meaning of the self-annihilation (3$\gamma _{\text{a}}$-annihilation) $^{%
\text{T}}$Ps rate $\lambda _{\text{T}}($obs.$)$ over the meaning calculated
in quantum elecrtodynamics ({\it QED})%
$$
\lambda _{\text{T}}\text{(theor.)}=7.03830\pm 0.00005\,(\sim 0.0007\%)\,\mu
s^{-1} 
$$
(`$\lambda _{\text{T}}$-anomaly' is revealed \cite{bib5}):%
$$
\frac{\lambda _{\text{T}}\text{(obs.) --- }\lambda _{\text{T}}\text{(theor.)}%
}{\lambda _{\text{T}}}=\text{(}0.19\text{[11]}\div 0.14\text{[12])}\pm \text{%
(0.02[11]}\div \text{0.023[12])\%.} 
$$

Coming back to features of lifetime spectra in neon, we shall notice that
the estimations ranges of the quantity $10^4\sim n\ll 10^5$ of nuclei
collective $^{22}$Ne, sustaining nuclear excitation $^{22*}$Ne$\,(2^{+})$ in a
final state of $\beta ^{+}$-decay $^{22}$Na, is made in \cite{bib3} on the basis of
a hypothesis about {\it NGR} of $\gamma _{\text{n}}$-quantum within {\it %
MCN/R/S} with taking into account insignificance of internal conversion
factor and within insignificance of pair conversion factor for transition $%
^{22*}$Ne$\,(2^{+})\rightarrow $ $^{22}$Ne$\,(0^{+})$. This estimation is
supported and specified in paper \cite{bib5} on the basis of other reasons%
$$
n=\,\stackrel{\_}{n}\cdot \eta =5.2780\cdot 10^4\cdot 0.09\simeq 0.5\cdot
10^4\eqno (1) 
$$

\noindent ($\overline{n}=5.2780\cdot 10^4\,$is the complete number of nuclei 
{\it MCNS}, and $\eta $ is the share of isotope $^{22}$Ne in natural neon),
developing an idea of $\beta ^{+}$-decay of nuclei like $^{22}$Na as
topological quantum transition:

--- about connection of the orthopositronium with `mirror Universe' \cite{bib13}, in
which from the point of view of the common observer the `antipode symmetry'
of energy and action is realized \cite{bib14}, and

--- about complete degeneration of {\it ortho-} and {\it parasuperpositronium}
in $N=2$ {\it supersymmetric} {\it QED} $(N=2$ {\it SQED}$)$ \cite{bib15}.

Experimental supervision of dependence of characteristic sites of lifetime
spectra on gaseus neon isotope composition \cite{bib16} (`isotope anomaly' \cite{bib5}) has
confirmed all preconditions of the phenomenology \cite{bib3}. Let's consider `shift' 
$\Delta $ (as a result of double supertransformation --- from fermion to
boson and back to fermion\footnote{%
It concerns to the positron in the orthopositronium, as electron represents
the so-called {\it entangled state} with others electrons in the observable
Universe.} --- a particle is transferred in other point of space \cite{bib17}) in
quality of a `lattice constant' of cellular, crystal like structure of
limited volume non-trivial space topology in a final state of $\beta ^{+}$%
-decay, and we identify it to the orthopositronium oscillation between our
Universe and `mirror Universe', which explains `$\lambda _{\text{T}}$%
-anomaly' by 2+1-splitting 3$\gamma _{\text{a}}$-annihilation of the
orthopositronium $^{\text{T}}$Ps$\rightarrow \gamma _{\text{a}}/2\gamma
^{\prime }$ into a single observable $\gamma _{\text{a}}$-quantum ($\sim
1.022$ MeV) and two not observable mirror quanta (with total energy $\sim
3.6\cdot 10^{-4}$ eV \cite{bib5}). Over last years the qualitative substantiation of
2+1-splitting 3$\gamma _{\text{a}}$-annihilation of the orthopositronium
with participation of the `mirror Universe' has been received by the method
of chronometric invariants (physical observable values) in generalized
4-dimensional space-time, the special case of which is the space-time of the
General Theory of Relativity \cite{bib18}.

The structure of {\it MCN/R/S} includes an insignificant share of nuclei of
atoms\linebreak $\stackrel{\_}{n}=5.2780\cdot 10^4$ from their complete number $\sim
5\cdot 10^{22}$ in volume of gas $V_{\text{g}}\sim \frac 43\pi R_{\text{g}%
}^3 $~cm$^3$ ($R_{\text{g}}\simeq 2$ cm is a radius of the measuring
chamber) at pressure $\sim 50\div 75$ atm. We receive an experimental
estimation of the {\it MCNRS} `lattice constant'

\vspace*{-3pt}
$$
\Delta _{\text{exp}}=\left( \frac{V_{\text{g}}}{\stackrel{\_}{n}}\right)
^{1/3}\simeq 8\cdot 10^{-2}\text{ cm.}\eqno (2) 
$$

\vspace*{-1pt}
\noindent This estimation is comparable with virtual fundamental length
(`shift')

\vspace*{-2pt}
$$
\Delta \simeq c\cdot \frac \hbar {\frac 37\Delta W}=\frac 4{\alpha ^4}\cdot
\frac \hbar {m_{\text{e}}c}\simeq 5.5\cdot 10^{-2}\text{ cm,}\eqno (3) 
$$

\noindent caused by exchange interaction, the so-called `annihilation
interaction' bringing to energy increase of the ortopositronium ground state
on size $\frac 37\Delta W\simeq 3.6\cdot 10^{-4}$ eV (`new force of the
annihilation' \cite{bib19}, $\Delta W$ is the hyperfine splitting of the ortho- and
parapositronium). In terms of physics it means, that the attraction between
electron and positron in $^{\text{T}}$Ps is weakened, because for a time $%
\sim \frac{\displaystyle\hbar }{\frac 37\displaystyle\Delta W}\simeq 2$ ps
they stay in the form of one virtual photon. This fact can be interpreted as
impossibility to locate $^{\text{T}}$Ps within the limits of volume, smaller
than $\Delta ^3$. Therefore radius {\it MCNS} $r_{\text{c}}$ we shall
determine from equality%
$$
\frac 43\pi r_{\text{c}}^3=\stackrel{\_}{n}\cdot \Delta ^3\text{, }r_{\text{c%
}}=1.28\text{ cm.}\eqno (4) 
$$

\vspace{3pt}
The formation of {\it MCN/R/S}, having macroscopic radius $r_{\text{c}%
}\simeq 1.28$ cm, occurs as the consequence of long-range action for baryon
charge: the nontrivial topology of limited `volume' of space-time in a final
state of $\beta ^{+}$-decay means occurrence in a lattice units the baryon
charge $B$ (its carrier is a proton --- $p$; the nuclei of substance atoms
interact with it by exchange), compensated by `hole' $\stackrel{\_}{B}(%
\stackrel{\_}{p})$ with negative mass in the `mirror Universe' (`antipode
symmetry' \cite{bib14}), with which substance does not interact.

For finding out coordination of the concept {\it MCN/R/S} we consider
initial phenomenological model \cite{bib3} in approximation of repeatedly resonant
scattering of $\gamma _{\text{n}}$-quantum with resonant cross section

\vspace*{-4pt}
$$
\sigma _{\text{r}}=f_{\text{M}}\cdot \frac{\lambda ^2(2I_1+1)}{2\pi (2I_0+1)}%
\text{ cm}^2\text{,} 
$$

\vspace*{-2pt}
\noindent where $f_{\text{M}}=\exp [-4\pi ^2\stackrel{\_\_}{x^2}$/ $\lambda
^2]$ is the M\"ossbauer factor ($\stackrel{\_\_}{x^2}$ is a root-mean-square
displacement of a nucleus $^{22}$Ne in a direction of $\gamma _{\text{n}}$%
-quantum), $\lambda =2\pi \hbar c$/$E_{\gamma _{\text{n}}}$ is length of a
wave of $\gamma _{\text{n}}$-quantum, $I_{\text{0}}$ and $I_{\text{1 }}$are
the spins of ground and exited states of $^{22}$Ne. As in {\it MCN/R/S} the
root-mean-square displacement of a nucleus has the radius of nuclear forces
action $\sim $(2--3)$\cdot 10^{-13}$ cm (`isotope anomaly' \cite{bib5}), and $\lambda
\sim 10^{-10}$ cm, then the M\"ossbauer factor gets the maximum value $f_{%
\text{M}}=1$. We should remind here, that from middle 1980s up to middle
1990s a hypothesis about stationary long-range action for baryon charge
(so-called `fifth force') was studied and is not confirmed by experiment,
but this does not affect non-stationary aspect of similar
non-electromagnetic long-range action in a final state of $\beta ^{+}$-decay.

Thus, resonant cross section $\sigma _{\text{r}}\simeq 7.5\cdot 10^{-21}$ cm$%
^2$. It is easy to receive an estimation of `length of resonant run' of $%
\gamma _{\text{n}}$-quantum between two consecutive acts of resonant
scattering on nuclei $^{22}$Ne in {\it MCNRS}, as $f_{\text{M}}=1$ (the
exact resonance) means, that length of run can be estimated as for gas of
elastic spheres%
$$
l_{\gamma _{\text{n}}}\sim \frac 1{\eta \nu \cdot \sigma _{\text{r}}}\simeq
\frac 1{0.09\cdot 50\cdot 2.7\cdot 10^{19}\cdot 7.5\cdot 10^{-21}}\simeq 1.1%
\text{ cm,}\eqno (5) 
$$

\noindent where $\eta \nu $ is a concentration of an isotope $^{22}$Ne in
natural neon at pressure \mbox{p~$\sim 50$~atm}. The received estimation seems to
exclude consecutive realization of the concept\linebreak {\it MCNRS}, because with
dense cell packing in structure of {\it MCNRS} the parameter of packing $%
1/12\simeq 0.083$ is close to a share of an isotope $^{22}$Ne in natural
neon ($\eta \simeq 0.09$), $i.e.$ distance between two nuclei $^{22}$Ne on
the way of $\gamma _{\text{n}}$-quantum equal to 2$\Delta \simeq 0.11$ cm,
and on the order less than $l_{\gamma _{\text{n}}}$. But it is necessary to
take into account the specific feature of kinetics of elementary processes
in {\it MCNRS}, instant participation in resonant scattering of $\gamma _{\text{n}%
} $-quantum $n=\stackrel{\_}{n}\eta \simeq 0.5\cdot 10^4$ nuclei $^{22}$Ne,
like in parallel chemical reactions, $i.e.$ macroscopic resonant cross
section grows $n$-multiple: $\Sigma _{\text{r}}=n\sigma _{\text{r}}$.

For the majority of the researchers, both theoretists and experimenters, the
problem of orthopositronium arose only in the second half of 1980s, after
numerous confirmations and specification \cite{bib10,bib11,bib12} of the first observation by
Michigan group of a discrepancy between theory and experiment \cite{bib9}. The next
decade passed in intensive search of the solution of the problem, but it was
not found. From middle 1990s, when it seemed that all experimental and
theoretical ideas were exhaused in these researches there came a pause. In
the last years the interest in this problem has revived again \cite{bib20,bib21}. But a
suggestion about single-quantum annihilation of $^{\text{T}}$Ps \cite{bib22} with
participation of the `mirror Universe' \cite{bib5,bib23} has not been investigated in
experiment.

The above analysis of `isotope anomaly' in neon allows to proceed to
estimation of an additional mode of single-quantum annihilation $^{\text{T}}$%
Ps. Let's consider the result of work \cite{bib24}, in which the probability of the $%
^{\text{T}}$Ps annihilation on one $\gamma _{\text{a}}$-quantum and neutral
supersymmetric gauge boson $U$ with spin 1 is calculated:%
$$
{\sf B}(^{\text{T}}\text{Ps}\rightarrow \gamma U)=3.5\cdot 10^{-8}\cdot
(1-x^4),\eqno (6) 
$$

\noindent where $x=(m_{\text{U}}/m_{\text{e}})\rightarrow 0$. An explanation
of `$\lambda _T$-anomaly' ($\sim 0.2\%$) lacks 4--5 orders. But it is already
clear, that in {\it MCNS} due to single-quantum virtual annihilation take
place the oscillations $^{\text{T}}$Ps$\rightleftharpoons $$^{\text{T}}$Ps$%
^{\prime }$ between our Universe and the `mirror Universe' (`there and
back', with shift $\Delta $ \cite[see footnote~($^1$)]{bib5}, 
{\it i.e.} occurs incoherent interaction $^{%
\text{T}}$Ps$\rightleftharpoons $$^{\text{T}}$Ps$^{\prime }$ with $\stackrel{%
\_}{n}$ `units' of space-like structures {\it MCNS}, and hence multiple
summation of probability (6). We receive missing 4--5 orders, and realistic
estimation $\sim 2\cdot 10^{-3}$ of the mode contribution

\vspace*{-8pt}
$$
^{\text{T}}\text{Ps}\,/^{\text{T}}\text{Ps}^{\prime }+M\rightarrow \gamma _{%
\text{a}}/2\gamma ^{\prime }+M\text{,}\eqno (7)\text{,} 
$$

\vspace*{2pt}
\noindent similar in all others attitudes to mode (6). Here mass $M$ in (7)
has a secondary origin: it is formed in gas on $\stackrel{\_}{n}$ units of 
{\it MCNS}, having an affinity to baryon charge of atomic nuclei.

Nevertheless there are some more difficulties: first, earlier in experiment
was received the estimation of the top limit of the probable contribution of
the $^{\text{T}}$Ps single-photon annihilation $\leq 4\cdot 10^{-4}\%$ \cite{bib25};
secondly, from the point of view of phenomenological model based on results
of supersymmetric consideration of the orthopositronium problem \cite{bib5,bib15,bib24}, a
single-photon mode of orthopositronium annihilation (three photon mixed $%
\gamma _{\text{a}}/2\gamma ^{\prime }$) does not agree with symmetry,
because photon propagates in a certain direction as the flat wave, {\it i.e.}
the central symmetry of space is broken.

The way out of this critical situation was prepared before discovery of
supersymmetry by definition of {\it notoph} \cite{bib26}: `contrary to mass-bearing
particles, helicity (projection of full angular momentum on a direction of
movement) is relativistic invariant for zero-mass particles. Under
discussion in a massless particle with zero helicity complementary to photon
according to its properties (helicity is equal to $\pm 1$) and therefore
referred to as `notoph'. In interactions notoph similar to photon bears spin
1. Notoph is described by antisymmetric tensor potential, while the field
strength is a 4-vector (instead of vector potential and tensor of
electromagnetic field strength) \dots when discussing new particles and their
interactions we should take notoph into account\dots '.

Helicity of notoph is equal to zero, {\it i.e.} it propagates as a spherical
wave. Thus, experimental task of search of single-photon annihilation $^{%
\text{T}}$Ps$\,/^{\text{T}}$Ps$^{\prime }\rightarrow \gamma _{\text{a}%
}/2\gamma ^{\prime }$ is transformed to search of single-notoph annihilation 
$^{\text{T}}$Ps$\,/^{\text{T}}$Ps$^{\prime }\rightarrow \gamma _{\text{a}%
}^{\circ }/2\gamma ^{\prime }$, and it changes the technique of experiment.
In measurements \cite{bib1,bib9,bib10,bib11,bib12,bib16,bib25} were fixed not only time and power
correlations of $\gamma $-quanta, but also spatial arrangement of detectors,
which breaks angular isotropy of measurements. For an establishment of the
contribution of supersymmetric single-notoph mode of $^{\text{T}}$%
Ps\thinspace $/^{\text{T}}$Ps$^{\prime }$ annihilation the technique
ensuring 4$\pi $-geometry of detecting notoph is needed. In work \cite{bib21} it is
suggested that the technique which was used in work \cite{bib27} should be improved
to achieve higher sensitivity. According to the authors of work \cite{bib21} for
realization of reliable measurements of the contribution of a mode $^{\text{T%
}}$Ps$\rightarrow ^{\text{T}}$Ps$^{\prime }$ in conditions of vacuum the
increase of 4$\pi $-calorimeters sensitivity is necessary.

In our suggestion which takes into account also `isotope anomaly' of
orthopositronium, it is necessary to allocate expected effect at $E_{\gamma
_{\text{a}}}\simeq 2m_{\text{e}}c^2$ on a background of peak of complete
annihilation energy $\sim 1.022$ MeV, instead of at $E_{\gamma _{\text{a}%
}}\simeq 0$, as in work \cite{bib27} and as it is offered in \cite{bib21}.

The offered here way of the ortopositronium problem decision summarizes
supervision, phenomenology and experimental results submitted in our
publication for 25 years, and is in essence complete substantiation of
decisive experiment for search single-notoph annihilation $^{\text{T}}$Ps$%
\,/^{\text{T}}$Ps$^{\prime }\rightarrow \gamma _{\text{a}}^{\circ }/2\gamma
^{\prime }$. It is based on the ideas which have arisen on `crossing' of
nuclear-physical directions and techniques of researches of a chemical
kinetics and a structure of substance --- the `{\it positronics}' and {\it %
gamma-resonance spectroscopy}, --- created in the beginning 60s in
Institute of Chemical Physics by V.I.~Goldanskii with co-workers. The author
is grateful to memory of the Academician Vitalii Iosifovich Goldanskii for
support of these publications, which stimulated statement of critical
experiment, for organizational support of experimental work \cite{bib16} and for the
shown interest to coming-to-be of this direction of the orthopositronium
dynamics researches.

The author thanks Prof. V.P.~Shantarovich for participation at an initial
stage of these researches.


{\small
\begin{thebibliography}{99}  {\parskip=-3pt
\bibitem{bib1}  Goldanskii \& Levin, Institute of Chemical Physics,
Moscow (1967), in Atomic Energy Review. Table of positron annihilation data,
edited by B.G.~Hogg and C.M.~Laidlaw and V.I.~Goldanskii and V.P.~%
Shantarovich, Vol.6 (IAEA, VIENNA) 1968, p.183.

\bibitem{bib2}  Levin B.M.\ and Shantarovich V.P., High Energy Chemistry, 11(4)
(1977) 382.

\bibitem{bib3}  Levin B.M.\ and Shantarovich V.P., Sov.\ J.\ Nucl.\ Phys., 34(6)
(1984) 1353.

\bibitem{bib4}  Pontecorvo B.M., Uspekhi Fizicheskikh Nauk, 141(4) (1987) 675
[in Russian].

\bibitem{bib5}  Levin B.M., Phys. Atomic Nuclei, 58 (1995) 380.

\bibitem{bib6}  Levin B.M., Gravitation and Cosmology, 5, Supplement (1999) 92.

\bibitem{bib7}  Zel'dovich Ya.B., Uspekhi Fizicheskikh Nauk, 123 (1977) 502 [in
Russian].

\bibitem{bib8}  Levin B.M., Sov.\ J.\ Nucl.\ Phys., 34(6) (1981) 1653.

\bibitem{bib9}  Gidley D.W., Rich A., Sweetman E.\ and West D., Phys.Rev.Lett.,
49 (1982) 525.

\bibitem{bib10}  Westbrook C.I., Gidley D.W., Conti R.S.\ and Rich A.,
Phys.Rev.Lett. 58 (1987) 1328.

\bibitem{bib11}  Westbrook C.I., Gidley D.W., Conti R.S.\ and Rich A., Phys.Rev.,
A40 (1989) 5489.

\bibitem{bib12}  Nico J.S., Gidley D.W., Rich A.\ and Zitzewitz P.W.,
Phys.Rev.Lett. 65 (1990) 1344.

\bibitem{bib13}  Glashow S.L., Phys.Lett., B167 (1987) 35.

\bibitem{bib14}  Linde A.D., Phys.Lett. B200 (1988) 272.

\bibitem{bib15}  Di Vecchia P.\ and Schuchhardt V., Phys.Lett., B155 (1985) 427.

\bibitem{bib16}  Levin B.M., Kochenda L.M., Markov A.A.\ and Shantarovich V.P.,
Sov.\ J.\ Nucl.\ Phys., 45(6) (1987) 1119.

\bibitem{bib17}  Freedman D.Z.\ and van Nieuwenhuizen P., Sci.Amer., 238(2)
(1978) 126.

\bibitem{bib18}  Borissova L.B.\ and Rabounski D.D. Field, vacuum, and the mirror
Universe (URSS, Moscow) 2001.

\bibitem{bib19}  Feynman R.P. Quantum electrodynamics (W.A.Benjamin, Inc., N.Y.)
1961.

\bibitem{bib20}  Adkins G.S., Melnikov K.\ and Yelkhovski A., Phys.Rev., A60
(1999) 3306.

\bibitem{bib21}  Foot R.\ and Gninenko S.N., Phys.Lett., B480 (2000) 171.

\bibitem{bib22}  Levin B.M.\ and Shantarovich V.P., Preprint (Institute of
Chemical Physics, Academy of Sciences USSR, Chernogolovka) 1985.

\bibitem{bib23}  Levin B.M., Sov.\ J.\ Nucl.\ Phys., 52(2) (1990) 342.

\bibitem{bib24}  Fayet P. and Mezard M., Phys.Lett. B104 (1981) 226.

\bibitem{bib25}  Gidley D.W., Nico J.S.\ and Skalsey M., Phys.Rev.Lett., 66 (1991)
1302.

\bibitem{bib26}  Ogievetskii V.I.\ and Polubarinov M.I., Sov.\ J.\ Nucl.\ Phys., 4
(1968) 216.

\bibitem{bib27}  Atoyan G.S., Gninenko S.N., Razin V.I.\ and Ryabov Yu.V.,
Phys.Lett., B220 (1989)~317.

}

\end{thebibliography}

}
\end{document}